\documentclass[useAMS,usenatbib,usegraphicx]{mn2e}
\usepackage{times}

\newcommand{\qso}{PKS 2126$-$158}
\newcommand{\kms}{\mbox{km\,s$^{-1}$}}
\newcommand{\kmsmpc}{\mbox{km\,s$^{-1}$\,Mpc$^{-1}$}}
\newcommand{\ergscm}{\mbox{erg\,s$^{-1}$\,cm$^{-2}$}}
\newcommand{\ergscmA}{\mbox{erg\,s$^{-1}$\,cm$^{-2}$\,\AA$^{-1}$}}
\newcommand{\lya}{Ly$\alpha$}
\newcommand{\mgi}{\hbox{Mg\,{\sc i}}}
\newcommand{\mgii}{\hbox{Mg\,{\sc ii}}}
\newcommand{\caii}{\hbox{Ca\,{\sc ii}}}
\newcommand{\civ}{\hbox{C\,{\sc iv}}}
\newcommand{\cii}{\hbox{C\,{\sc ii}}}
\newcommand{\siii}{\hbox{Si\,{\sc ii}}}
\newcommand{\siiii}{\hbox{Si\,{\sc iii}}}
\newcommand{\siiv}{\hbox{Si\,{\sc iv}}}
\newcommand{\oi}{\hbox{O\,{\sc i}}}
\newcommand{\oii}{\hbox{[O\,{\sc ii}]}}
\newcommand{\oiii}{\hbox{[O\,{\sc iii}]}}
\newcommand{\alii}{\hbox{Al\,{\sc ii}}}
\newcommand{\aliii}{\hbox{Al\,{\sc iii}}}
\newcommand{\feii}{\hbox{Fe\,{\sc ii}}}
\newcommand{\nv}{\hbox{N\,{\sc v}}}
\newcommand{\hbeta}{H$\beta$}
\newcommand{\halpha}{H$\alpha$}
\newcommand{\ie}{i.e.\ }
\newcommand{\eg}{e.g.\ }

\title[Discovery of a $z=0.66$ galaxy group]{Multi-object spectroscopy of the field surrounding \qso: Discovery of a $\bmath{z=0.66}$ galaxy group}
\author[M. T. Whiting et al]{Matthew T. Whiting$^{1,2}$\thanks{E-mail:
  Matthew.Whiting@csiro.au} and Rachel L. Webster$^{3}$ and 
  Paul J. Francis$^{4}$\\
$^{1}$School of Physics, University of New South Wales, Sydney, NSW,
  2052, Australia\\
$^{2}$Australia Telescope National Facility, P.O.\
  Box 76, Epping, NSW, 1710, Australia\\
$^{3}$School of Physics, University of Melbourne, VIC, 3010,
  Australia\\
$^{4}$Research School of Astronomy and Astrophysics, Australian
  National University, ACT, 0200, Australia}

\begin{document}
\maketitle

\label{firstpage}

\begin{abstract}
The high-redshift radio-loud quasar \qso\ is found to have a large
number of red galaxies in close apparent proximity. We use the Gemini
Multi-Object Spectrograph (GMOS) on Gemini South to obtain optical
spectra for a large fraction of these sources. We show that there is a
group of galaxies at $z\sim0.66$, coincident with a metal-line
absorption system seen in the quasar's optical spectrum. The
multiplexing capabilities of GMOS also allow us to measure redshifts
of many foreground galaxies in the field surrounding the quasar.

The galaxy group has five confirmed members, and a further four
fainter galaxies are possibly associated. All confirmed members
exhibit early-type galaxy spectra, a rare situation for a \mgii\
absorbing system. We discuss the relationship of this group to the
absorbing gas, and the possibility of gravitational lensing of the
quasar due to the intervening galaxies.
\end{abstract}


\section{Introduction}
\label{sec-intro}

Quasars located at high redshifts provide excellent probes of the
intervening universe. As their light travels through the universe, it
is partially absorbed by neutral hydrogen or metals such as magnesium
-- an effect detectable in the quasars' optical and UV spectra. These
absorption lines provide an effective way of probing the evolution of
metal-bearing gas over a large range of redshifts, as the sensitivity
to absorption is largely independent of redshift, depending only on
the background source's surface brightness.

The \mgii\ absorption systems in particular are most likely associated
with foreground galaxies \citep{bergeron91,steidel94,steidel02}, based
on the velocity structure within the systems, the clustering
properties of the absorbers, and the presence of metals that have most
likely been produced locally. When identified, most of these galaxies
appear to be spirals \citep[\eg][]{steidel02}, although absorbers have
been found that span the range from late-type spirals to galaxies
resembling present-day ellipticals \citep{steidel94}. The
observational challenge then is to correctly identify the absorbing
galaxies and build up a picture of the absorbing system. Such
identifications allow one to track the evolution of metal-bearing gas
and its relationship to galactic environment.

The identification of the object responsible for an absorption system
involves finding a galaxy close to the line-of-sight to the quasar
that has a redshift matching that of the absorption system. As such
galaxies are often faint, to examine many in the vicinity of a given
quasar requires relatively large amounts of telescope time, or a large
degree of multiplexing. The advent of multi-object spectrographs on
large telescopes has enabled more systems and galaxies to be examined,
greatly assisting the identification of absorbing systems.

The identification of nearby galaxies close on the sky to a
high-redshift quasar also allows us to address the question of
gravitational lensing. The high redshifts of these quasars on face
value imply very large luminosities, placing them at the high end of
the luminosity function, where the slope of the function is
steepest. If a small amount of magnification due to gravitational
lensing is taking place, the intrinsic luminosity of the quasar will
be reduced, changing the shape of the luminosity function. This is
particularly important for the highest redshift quasars (such as those
at $z>6$, \eg \citet{richards04}), but also for those around $z\sim
3-4$.

In this paper we address these issues by targeting \qso, a luminous
high-redshift quasar that exhibits a number of metal-line absorption
systems and has a large number of galaxies close to its line of sight,
using the multiplexing capabilities of GMOS on the Gemini South
telescope. We use imaging and low-resolution spectroscopy to obtain
redshifts for many of the galaxies in the field, focussing on the
$z<1$ environment. 

We describe this quasar and our observations of it and the surrounding
field in Section~\ref{sec-obs}, while the results of the Gemini/GMOS
observations are presented in Section~\ref{sec-results}. The
implications of these results for the absorption line systems seen in
the quasar's spectrum are discussed in Section~\ref{sec-absn}, and the
possibility of magnification of this quasar due to gravitational
lensing is discussed in Section~\ref{sec-lens}. A summary of the
results is found in Section~\ref{sec-summary}. Note that throughout
this paper we assume a standard $\Lambda$CDM cosmology, with
$H_0=71\,\kmsmpc$, $\Omega_m=0.27$ and $\Omega_\Lambda=0.73$
\citep{spergel03}.

\section{Target Field and Observations}
\label{sec-obs}

\subsection{The quasar \qso}

The quasar \qso\ is a radio-loud, flat-spectrum quasar at a redshift
of $z=3.2663$. It was first identified by \citet*{condon77} in optical
follow-up of flat-spectrum sources from the Parkes 2700 MHz surveys,
and at the time its redshift was measured it was only the fifth radio
quasar with $z>3$ \citep{jauncey78:2126}.

\qso\ is known to be a very bright object at radio ($S_{2.7\,{\rm
GHz}}=1.17\, \mbox{Jy}$, \citet{pkscat90}), optical/near-infrared
($V=16.92, H=14.89$, \citet*{francis00}) and X-ray ($F_{0.1-2.4\, {\rm
keV}} = 2\times10^{-12}\, \ergscm$, \citep{siebert98})
frequencies. These fluxes, combined with its relatively high redshift,
mean it is among the most luminous quasars known.

The optical/near-infrared spectrum of \qso\ has a blue power-law shape
for wavelengths longer than $V$ band \citep{francis00}, where the
\lya\ line falls. The flux drops off at shorter wavelengths due to
absorption by the intervening gas of the Lyman alpha
forest. Absorption systems with strong metal lines are observed at the
redshifts indicated in Table~\ref{tab-lines}. The relatively strong
\mgii\ system at $z=0.663$ is particularly relevant for this paper,
and is discussed in more detail in Section~\ref{sec-absSystem}.

The peculiar concentration of objects in the vicinity of \qso\ was first
noticed from $K_n$-band images taken on 6 June 1993 by
\citet{drinkwater97} with IRIS \citep{allen93:IRIS} on the
Anglo-Australian Telescope in the course of identification of sources
in the Parkes Half-Jansky Flat-spectrum Sample (PHFS). 
There are $\sim16$ sources visible within 30 arcsec of the quasar,
principally toward the east and south.

\citet{veron90} obtained spectra of the two bright sources to the west
(designated therein as C1 and C2, at distances of $\sim5$ and $\sim10$
arcsec west of the quasar respectively). They were able to measure the
redshift of C2 (an emission line galaxy) as $z=0.210$, while the
spectrum of C1 was inconclusive. This is the only redshift information
available from the literature for objects in the vicinity of \qso.

\begin{table}
\caption{Metal-line systems observed in the spectrum of \qso\
  \citep{d'odorico98,giallongo93}.}
\label{tab-lines}
\begin{tabular}{lll} \hline\hline
$z_{\rm abs}$ &Velocity range &Metal lines\\ \hline
0.6631 &$\delta v \sim 215\,\kms$ &\mgii, \mgi, \caii \\
2.3313 &&\civ, \siii\\
2.3941 &$\delta v \sim 180\,\kms$ &\civ, \siiv, \siii, \alii\\
2.4597 &&\civ, \feii \\
2.5537 &&\civ\\
2.6378 &$\delta v \sim 286\,\kms$ &\civ, \cii, \alii, \aliii, \siii, \siiii, \\
       &&~\siiv, \mgi, \oi, \feii\\
2.6788 &&\civ, \siiv, \cii, \feii, \nv\\
2.7281 &&\civ, \siiv, \siii, \feii\\
2.7689 &$\delta v \sim 350\,\kms$ &\civ, \alii, \siii, \cii, \oi, \siiv, \\
       &&~\aliii, \feii\\
2.8195 &&\civ, \cii, \siii, \oi\\
2.9071 &&\civ, \alii, \siiii, \siiv\\
2.9675 &&Lyman limit system\\
3.2165 &&\civ, \siiv, \siii \\ 
\hline
\end{tabular}
\end{table}

\subsection{Gemini Observations}

\begin{figure*}
\begin{minipage}{170mm}
  \includegraphics{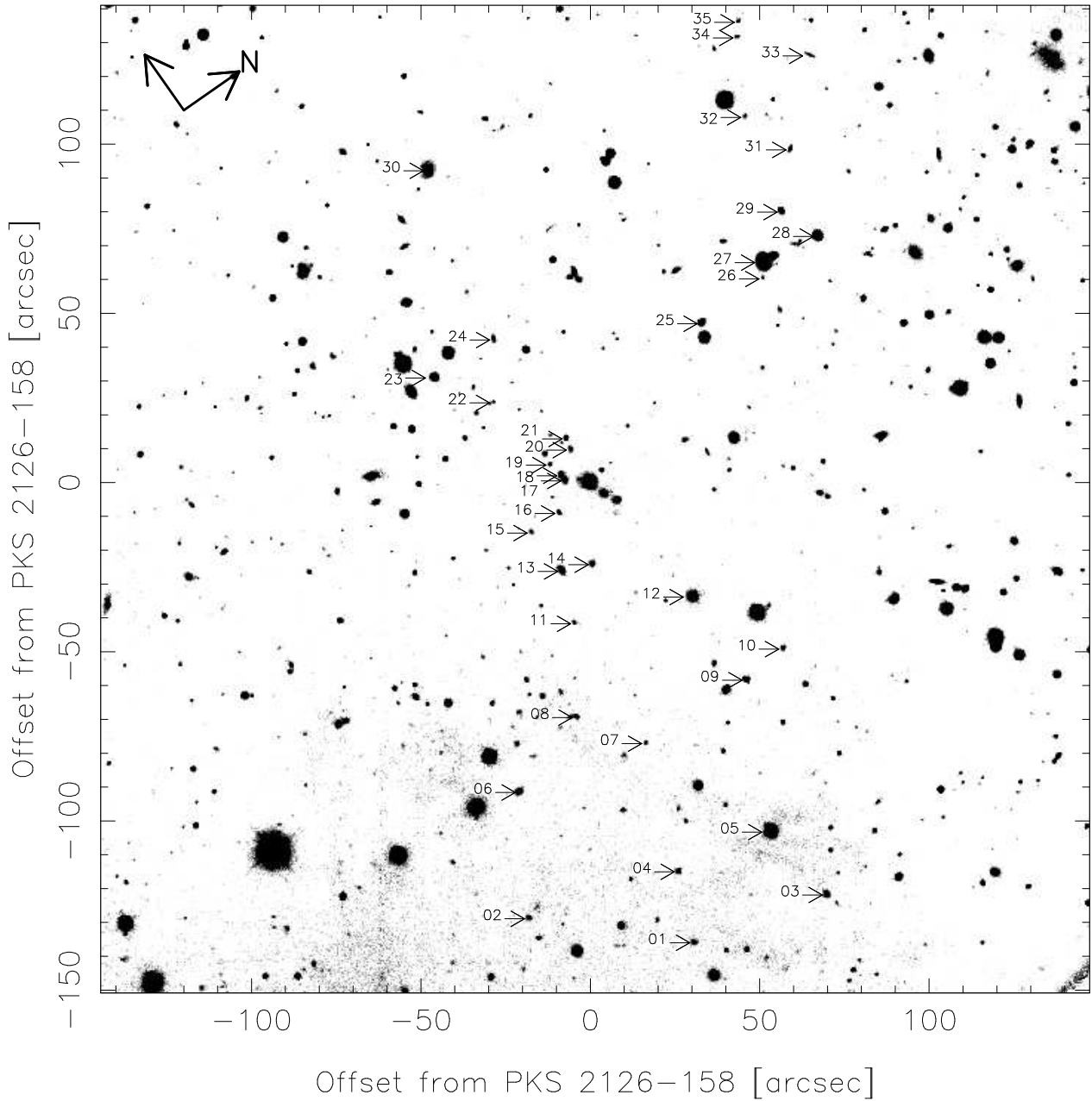}
  \caption{An image of the field taken in $i'$ band with
  GMOS-South. The image has been cropped to a square region containing
  all spectroscopic targets. The locations of each of the targets are
  indicated with their identification numbers used herein. The
  orientation of the field is indicated by the arrows at top-left, and
  offsets are measured with respect to the quasar \qso.}
  \label{fig-gmos}
\end{minipage}
\end{figure*}

We used the Gemini Multi-Object Spectrograph
\citep[GMOS,][]{hook02:GMOS} on the Gemini South telescope, in
Multi-Object Spectroscopy (MOS) mode, to obtain spectra for as many of
the nearby objects as possible. Pre-imaging of the field was done with
GMOS-South on 2003 July 15. The image, shown in Fig.~\ref{fig-gmos},
was constructed from six 300\,sec exposures in the $i'$ filter (Gemini
filter i\_G0327), and has image quality of 0.7~arcsec (the image
was binned $2\times2$ on-chip).

\begin{figure*}
\begin{minipage}{170mm}
  \includegraphics{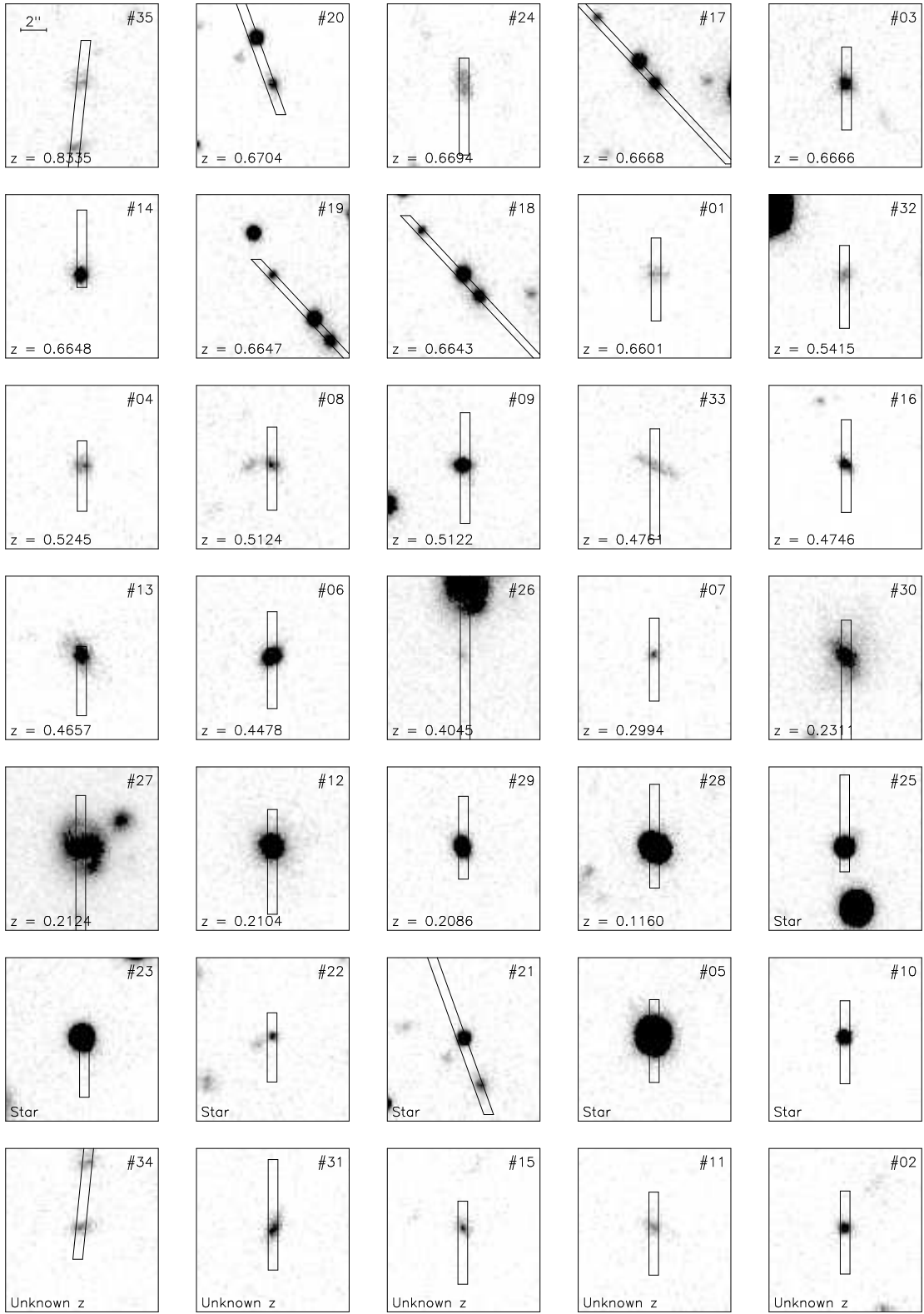}
\end{minipage}
\caption{GMOS $i'$ images of individual targets, listed in the same
  order as in Table~\ref{tab-results}. The location of the
  slit used is indicated. Each image is centred on the object
  indicated at the top-right. The angular scale, which is the same in
  each image, is shown in the image of \#35, at top-left.}
\label{fig-images}
\end{figure*}

Since GMOS uses multiple slits in its multi-object mode, one cannot
have slits overlapping in the dispersion direction (the horizontal
direction in Fig.~\ref{fig-gmos}), particularly when the target
objects are close together on the sky. For this project, this places a
limit on the number of sources that can be observed in a single
pointing (we had time for just a single mask setting). Fortunately, a
judicious choice of position angle (in this case, 215\degr\ East of
North) can enable a large fraction of the sources of interest to be
placed on a slit. The large field of view of GMOS ($\sim5.5$~arcmin)
also enabled a number of potentially interesting sources further from
the quasar to be placed under a slit. These were selected either on
morphology, or on their location in the field (\ie whether a slit
could be placed over them based on previously allocated slits). The
locations of the targeted objects are shown in Fig.~\ref{fig-gmos},
while Fig.~\ref{fig-images} shows each object in more detail, as well
as the location of each slit.

Note that the limiting magnitude of the image is approximately
$i'=24.6$. At this magnitude, an $L_\ast$ galaxy\footnote{We assume
$M_i^\ast=5\log_{10}h-21.00$ \citep{bell03}, and calculate
K-corrections with template SEDs from \citet{poggianti97}} is visible
out to a redshift of 5. However, at the magnitude of the faintest
spectroscopic target ($i'=23.33$), an $L_\ast$ galaxy can be seen only
out to $z=0.73$ (although a $3L_\ast$ galaxy survives the K-correction
and can be seen out to $z\sim5$).

The field was observed in queue mode on two nights -- 2003 September
24 and 25 -- with the chosen slit configuration, using the R150 grating
and the \verb|GG455| long-pass filter, for a total exposure time of 2
hours. This was divided into sets of $3\times 1200\,\mbox{sec}$
exposures at each of two grating settings, one on each of the
nights. The data was binned on-chip by a factor of 2 in both spatial
and dispersion directions.

The image quality, as measured from alignment images (with the mask in
place) taken prior to the spectral observations, was noticeably worse
on the second night (0.5\,arcsec on night one, 0.95\,arcsec on night
two). This limited the signal-to-noise achieved in the spectra, and
consequently several sources were unable to have redshifts measured (the
spectra where redshifts were measured typically had S/N $\ga5$, while
these few were significantly less than that).

The data were reduced using standard procedures from the Gemini {\sc
IRAF} package. The spectra were bias-subtracted and flat-fielded using
calibration frames from the Gemini GCAL unit. Wavelength calibration
using CuAr arc spectra, as well as sky-subtraction, were performed on
an individual slit basis, and the spectra were flux-calibrated using
observations of the standard star EG131.

\section{Results}
\label{sec-results}

\begin{table*}
\begin{minipage}{170mm}
\caption{ Observational data for galaxies in the field of \qso,
  numbered according to Fig.~\ref{fig-gmos}.  The $i'$ magnitude is
  measured from the GMOS pre-imaging data. The positional offsets
  $\Delta$RA, $\Delta$Dec and $\Delta\theta$ are the angular distances
  from \qso, while the impact parameter $\rho$ is the linear distance
  from the galaxy to the line-of-sight to \qso. For the comments on
  the spectra: PEG = Passive Elliptical Galaxy; ELG = Emission Line
  Galaxy; AGN = Active Galactic Nucleus.  }
\label{tab-results}
\begin{tabular}{@{}llllrrrrl@{}}\hline\hline
Object &$i'$ &$M_{i'}$ &$z$ &$\Delta$ RA &$\Delta$ Dec &$\Delta\theta$ &$\rho$ &Comments\\
 &[mag] &[mag] & &[arcsec] &[arcsec] &[arcsec] &[kpc] &\\\hline
35         &22.46 &-21.33 &0.8335 &  42.1 & 136.7 & 143.0 &1090.7 &Emission line at 6830\AA, ID as \oii \\
20         &21.82 &-21.58 &0.6704 &  10.2 &   4.6 &  11.2 &  78.5 &PEG \\
24         &21.48 &-21.92 &0.6694 &  47.4 &  18.3 &  50.8 & 356.3 &PEG + strong \oii \\
17         &20.79 &-22.60 &0.6668 &   6.3 &  -3.7 &   7.3 &  51.2 &PEG \\
03         &20.91 &-22.48 &0.6666 &-127.2 & -59.7 & 140.5 & 983.3 &PEG + weak \oii \\
14         &20.88 &-22.51 &0.6648 & -14.5 & -19.4 &  24.2 & 169.3 &PEG \\
19         &22.08 &-21.30 &0.6647 &  12.6 &  -2.4 &  12.8 &  89.7 &PEG \\
18         &20.34 &-23.04 &0.6643 &   8.2 &  -3.2 &   8.8 &  61.2 &PEG \\
01         &22.74 &-20.63 &0.6601 &-103.2 & -93.8 & 139.4 & 971.9 &PEG + weak \oii \\
32         &22.11 &-20.75 &0.5415 &  24.4 & 114.8 & 117.3 & 744.4 &\oii, with weak \hbeta\ and \oiii  \\
04         &21.88 &-20.89 &0.5245 & -87.4 & -79.2 & 118.0 & 736.0 &Strong, extended \oii\ emission \\
08         &21.71 &-21.00 &0.5124 & -36.5 & -59.1 &  69.5 & 428.5 &PEG + \oii \\
09         &20.43 &-22.28 &0.5122 & -71.3 & -21.3 &  74.4 & 458.7 &PEG \\
33         &21.88 &-20.63 &0.4761 &  19.3 & 140.6 & 141.9 & 839.9 &\oii \\
16         &21.60 &-20.90 &0.4746 &   2.2 & -12.7 &  12.8 &  75.9 &\oii \\
13         &20.38 &-22.08 &0.4657 &  -8.3 & -26.2 &  27.5 & 160.7 &Starburst (higher-order Balmer lines + \oii) \\
06         &20.02 &-22.33 &0.4478 & -35.5 & -86.9 &  93.9 & 536.8 &ELG (\hbeta, \oiii, \oii) \\
26         &23.33 &-18.74 &0.4045 &  -7.4 &  78.7 &  79.0 & 425.3 &ELG (\hbeta, \oiii, \oii) \\
07         &22.75 &-18.51 &0.2994 & -57.8 & -53.7 &  78.9 & 348.3 &ELG (\hbeta, \oiii, \oii) \\
30         &19.21 &-21.35 &0.2311 &  92.1 &  48.1 & 103.9 & 379.8 &PEG + weak \hbeta\ \& \oiii \\
27         &18.14 &-22.19 &0.2124 &  -4.8 &  82.8 &  82.9 & 284.0 &Barred spiral \\
12         &19.15 &-21.15 &0.2104 & -44.2 & -10.3 &  45.4 & 154.4 &PEG \\
29         &19.98 &-20.30 &0.2086 &  -0.4 &  98.0 &  98.0 & 331.1 &PEG + weak AGN?  \\
28         &17.95 &-20.84 &0.1160 & -13.3 &  98.2 &  99.1 & 205.8 &PEG \\
25         &18.87 &--- &0.0000 &  -0.0 &  57.4 &  57.4 &--- &Star \\
23         &17.98 &--- &0.0000 &  55.3 &  -0.9 &  55.3 &--- &Star \\
22         &22.17 &--- &0.0000 &  36.8 &   3.1 &  36.9 &--- &Star \\
21         &20.49 &--- &0.0000 &  13.2 &   6.6 &  14.7 &--- &Star \\
05         &16.44 &--- &0.0000 &-103.1 & -53.9 & 116.3 &--- &Star \\
10         &20.66 &--- &0.0000 & -74.9 &  -7.6 &  75.3 &--- &Star \\
34         &22.14 &--- &--- &  39.8 & 132.7 & 138.5 &--- &No redshift measured \\
31         &21.42 &--- &--- &   8.0 & 114.5 & 114.8 &--- &No redshift measured \\
15         &22.29 &--- &--- &   5.6 & -22.1 &  22.8 &--- &No redshift measured \\
11         &22.58 &--- &--- & -20.3 & -36.7 &  41.9 &--- &No redshift measured \\
02         &21.61 &--- &--- & -59.2 &-116.0 & 130.2 &--- &No redshift measured \\
\hline
\end{tabular}
\end{minipage}
\end{table*}

The measured redshifts for galaxies in the field of \qso\ are shown in
Table~\ref{tab-results}. The galaxies are numbered according to their
position on the image, and sorted in the table in order of decreasing
redshift. The $i'$ magnitudes shown are measured from the pre-imaging
data (\ie the image shown in Fig~\ref{fig-gmos}), calibrated on the
Landolt standard fields SA95 and TPhe, using the transformations from
\citet{fukugita96}. We have converted these magnitudes into absolute
magnitudes, $M_{i'}$. The K-corrections were calculated using the
template SEDs from \citet{poggianti97} and the filter transmission
function for filter i\_G0302 provided on the Gemini web
site.\footnote{Note that this transmission function is for the
equivalent $i'$ filter at Gemini North, believed to be similar to the
one used at Gemini South. The transmission function for the Gemini
South filter was not available at time of writing.} 

\begin{figure*}
\begin{minipage}{170mm}
  \includegraphics{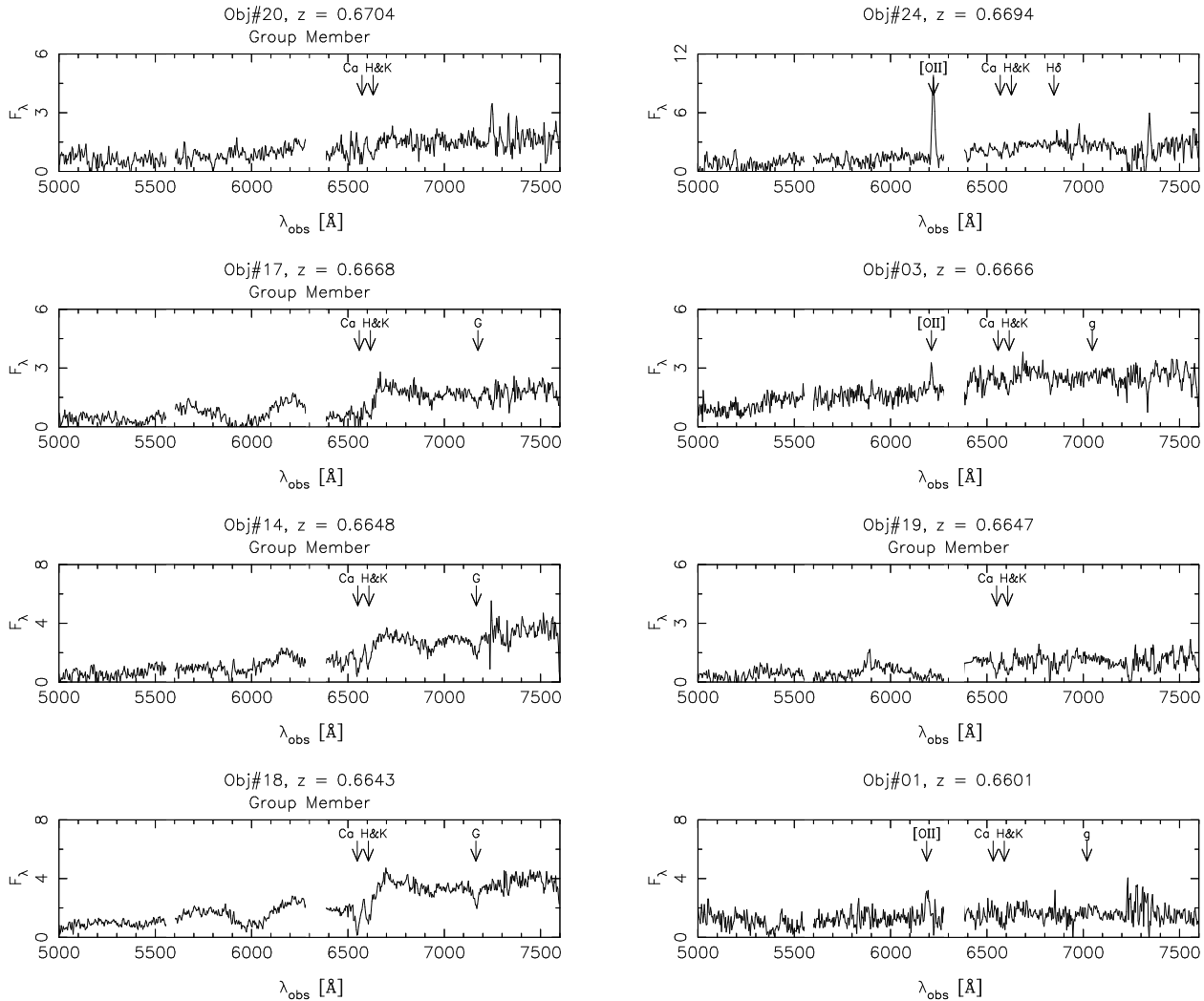}
\end{minipage}
\caption{Reduced GMOS-South spectra of the eight galaxies at redshifts
  of $z\sim0.66$, listed in order of decreasing redshift. Likely
  members of the group are indicated. The flux scale is $F_\lambda\
  [10^{-18}\ergscmA]$, plotted as a function of observed
  wavelength. Note that the varying flux scale from spectrum to
  spectrum. The identities of certain spectral features are indicated,
  and residuals from night-sky emission lines at 5577\AA\ and
  6300/6363\AA\ have been masked out. The noticeable spike at $\sim
  5890$\AA\ in the spectrum of Obj\#19 is a sky-subtraction artifact,
  caused by the location of this object close to the edge of a tilted
  slit (see the image in Fig.~\ref{fig-images}).}
\label{fig-spectra}
\end{figure*}

We find 8 galaxies with redshifts in the range
$0.66<z<0.67$, straddling the low redshift metal-line absorption
systems seen by \citet{d'odorico98}. The spectra of these galaxies are
shown in Fig.~\ref{fig-spectra}.

\begin{figure*}
\begin{minipage}{170mm}
  \includegraphics{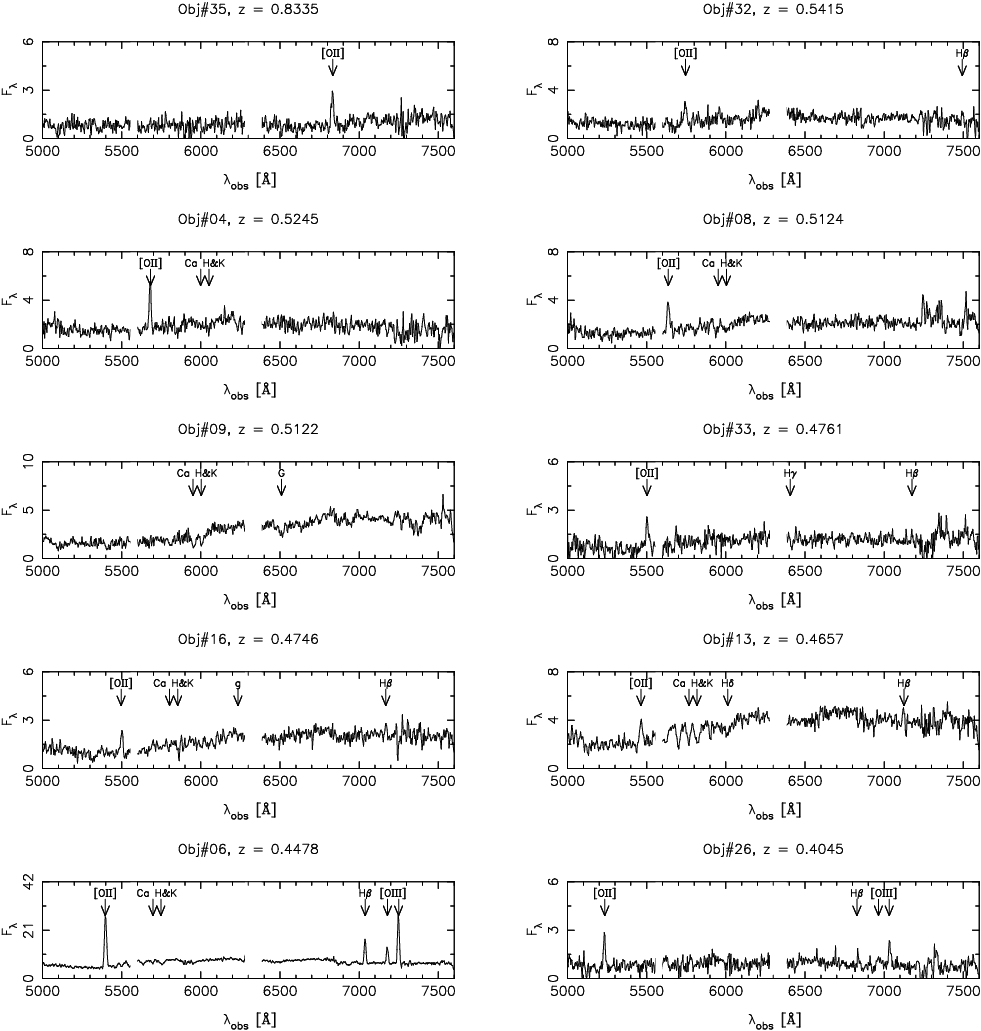}
\end{minipage}
\caption{Reduced GMOS-South spectra of confirmed galaxies in the field
  of \qso\ with $z>0.4$, listed in order of decreasing redshift. The
  flux scale is $F_\lambda\ [10^{-18}\ergscmA]$, plotted as a function
  of observed wavelength. The identities of certain spectral features
  are indicated, and residuals from night-sky emission lines at
  5577\AA\ and 6300/6363\AA\ have been masked out.}
\label{fig-spectra-hi}
\end{figure*}

\begin{figure*}
\begin{minipage}{170mm}
  \includegraphics{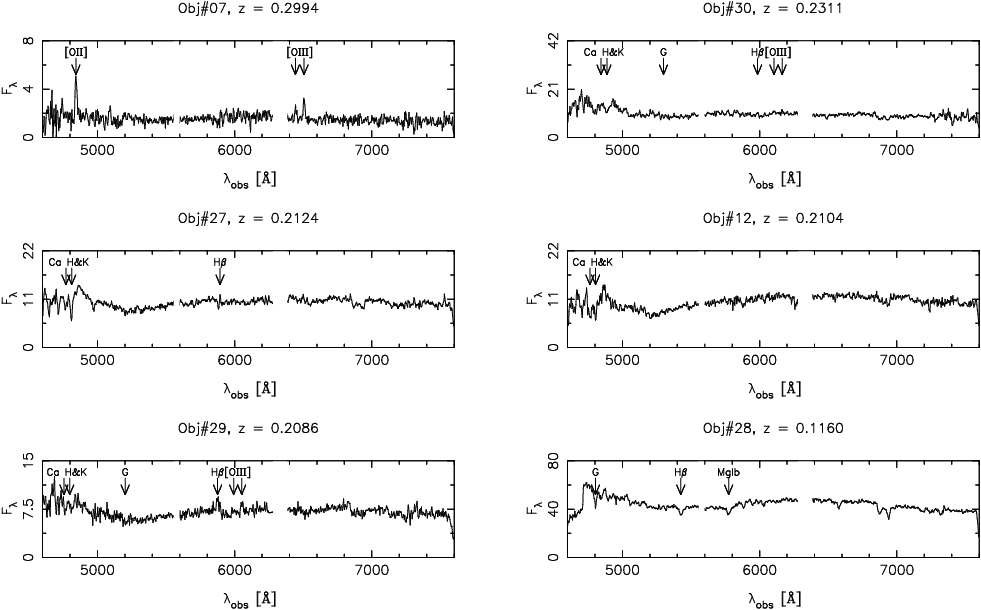}
\end{minipage}
\caption{Reduced GMOS-South spectra of confirmed galaxies in the field
  of \qso\ with $z<0.4$, listed in order of decreasing redshift. The
  flux scale is $F_\lambda\ [10^{-18}\ergscmA]$, plotted as a function
  of observed wavelength. The identities of certain spectral features
  are indicated, and residuals from night-sky emission lines at
  5577\AA\ and 6300/6363\AA\ have been masked out.}
\label{fig-spectra-lo}
\end{figure*}

\begin{figure}
  \includegraphics{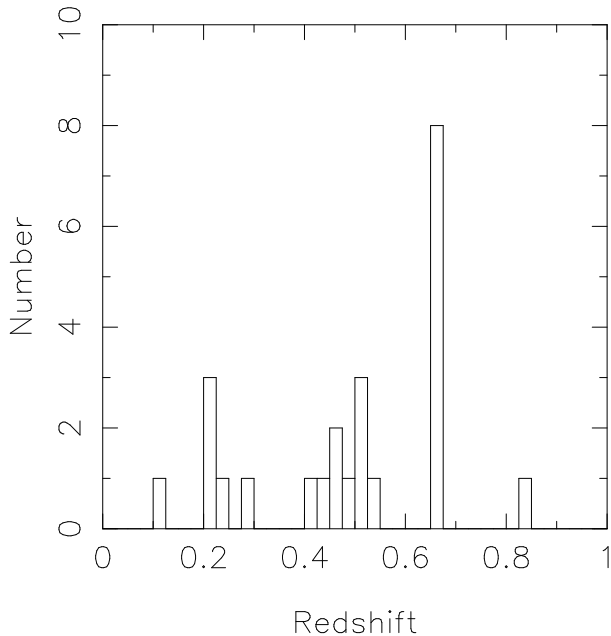}
  \caption{Distribution of measured redshifts of galaxies in the field
  of \qso.}
  \label{fig-zdist}
\end{figure}

The velocity dispersion of these eight galaxies is $\sigma=430\,\kms$,
centred on a redshift of $z_g=0.6660$. Three of these galaxies,
however, are at large spatial spearations from those close to the
quasar. If we include just the four closest to the quasar -- \#17,
\#18, \#19 and \#20 -- as well as \#14, which is just close enough to
be considered associated, we get $\sigma=355\,\kms$ and $z_g=0.6662$.
This velocity dispersion is characteristic of a group environment,
rather than a more dense cluster environment which would show $\sigma
\sim 500-1200\,\kms$. It is, in fact, a relatively poor group, with
only five members brighter than $M_{i'}=-21$. There are at least four
fainter galaxies for which we have not obtained spectra that are also
candidates for inclusion in the group (see Sec.~\ref{sec-absn} for
further discussion). The remaining galaxies at $z\sim0.66$ (\#24, \#03
and \#01) are part of the same large-scale structure that the group is
embedded in, but not necessarily directly associated with the group
itself.

In addition to this structure, we find several other features in
redshift-space. There are two other main structures seen in the field,
at $z\sim0.21$ and $z\sim0.46$ (and possibly $z\sim0.51$), although
these are not as tightly grouped on the sky as the main group at
$z\sim0.66$. There are a number of field galaxies at other redshifts
as well. We are thus seeing several features in the large-scale
structure along the line-of-sight to \qso.  The spectra of the
remaining galaxies with measured redshifts are shown in
Figs.~\ref{fig-spectra-hi}~\&~\ref{fig-spectra-lo}, and the
distribution of galaxy redshifts can be seen in Fig~\ref{fig-zdist}.

Of the remainder, six objects were identified as stars, and a further
five had spectra of too low quality for a redshift measurement. Of
these latter objects, two (\#11 and \#15) have extended morphologies
in the $i'$ GMOS image, and $i'-K$ colours redder than all the
confirmed group members ($i'-K\sim 4.7$ and $4.0$ respectively, cf.\
$i'-K\sim3$ for the group members). They are thus either red galaxies
in the $z\sim0.66$ group (more likely in the case of \#15), or
background galaxies (and redder due to their greater redshift).

\section{Association with absorbing system}
\label{sec-absn}

\subsection{Is one of the identified galaxies the absorber?}
\label{sec-absSystem}

As seen in Section~\ref{sec-results}, we find a concentration of
galaxies in the redshift range $0.66<z<0.67$. This coincides in
redshift-space with a metal-line absorption system detected by
\citet{giallongo93} and \citet{d'odorico98}. The system consists of
four components in velocity, with three species detected -- \mgii\ and
\caii, plus \mgi\ in the two highest column density components. The
redshifts of the components are $z=0.6625$, 0.6628, 0.6634 and 0.6637
\citep{d'odorico98}. This system is quite strong: the two strongest
\mgii\ components (at $z=0.6628$ and $z=0.6625$) have column densities
of $4.1\times10^{17}$ cm$^{-2}$ and $5.8\times10^{16}$ cm$^{-2}$
respectively \citep{ryabinkov03,d'odorico98}. Using the Doppler
parameters quoted by \citet{d'odorico98}, we can convert these to 
equivalent widths of 0.44\AA\ and 0.41\AA\ respectively (for the \mgii\
$\lambda2798$ line in the rest frame).

A key question is whether this system can be associated with the group
of galaxies detected in our GMOS observations. The nearest of the
galaxies to \qso\ is Galaxy \#17, which lies at a distance of
51.2\,kpc to the quasar's line-of-sight (hereinafter QLOS). This
places the QLOS in the outer parts of the galaxy's halo. This
separation is comparable to those seen in other absorption systems
(see, for example, \citet{steidel97}). This galaxy, however, does show
a large velocity offset from the absorbing system, with velocity
differences $\Delta v \equiv v_{\rm gal}-v_{\rm abs}$ with respect to
the four components ranging from $430-610\,\kms$. Galaxy \#18 provides
a much closer match to the velocities ($\Delta v = 74-253\,\kms$), and
is only 61.2\,kpc from the QLOS. It is also the brightest galaxy in
the group, with $i'=20.34$.

A relationship has been noted by \citet{lanzetta90} between the
equivalent width $W$ of the \mgii\ absorption system and the impact
parameter $\rho$ of the associated galaxy, where $W \propto
\rho^{-0.92 \pm 0.16}$. By converting the column density and Doppler
parameters of the absorption lines, we can obtain the equivalent
widths for the system: the four \mgii\ $\lambda2796$ lines from
\citet{d'odorico98} yield rest-frame values of $W=0.41, 0.44, 0.13$
and 0.13\AA\ (in order of increasing redshift).

Comparing with the \citet{lanzetta90} relationship, we find that the
two stronger systems have $W$ values corresponding to $\rho\sim30$~kpc,
while the two weaker systems correspond to $\rho\sim110$~kpc. The
galaxies we have observed to be in the group all lie at impact
parameters between these two values. There is, however, an object
closer than the closest observed galaxy to the QLOS (\# 17), which,
should it prove to be at the same redshift, may be a counterpart to
one of the stronger systems (see Section~\ref{sec-altgal} for more
discussion on this source).

If we focus just on the confirmed group members, the
\citet{lanzetta90} relationship taken at face value indicates that
none of these galaxies are suitable counterparts for the absorption
system(s). Recent studies, however, such as that presented in
\citet{churchill05}, indicate that there is much more scatter in the
$W-\rho$ relationship than the \citet{lanzetta90} relationship would
suggest. This calls into doubt its utility in making {\it predictions}
about the likelihood of potential absorbers. The \citet{churchill05}
results suggest that the observed galaxies can indeed be considered
potential hosts of the absorber, even with their relatively large
impact parameters. The identity of the absorber then is still an open
question. 

\subsection{An alternative absorbing galaxy.}
\label{sec-altgal}

\begin{figure}
  \includegraphics{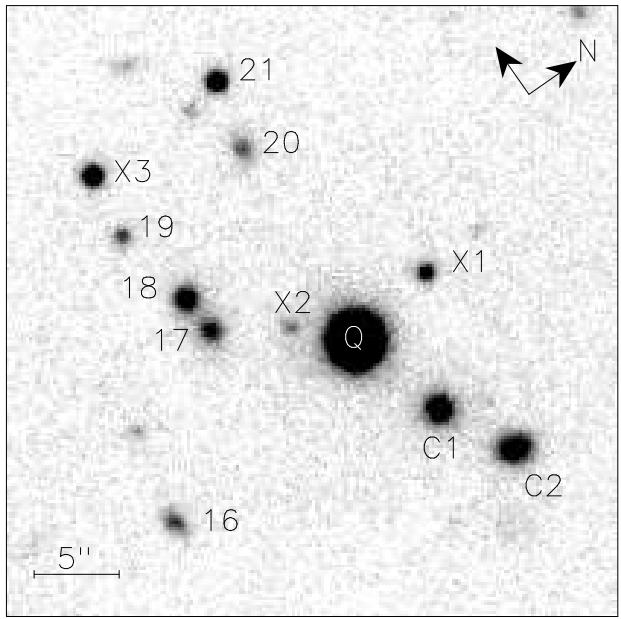}
  \caption{A $30\arcsec\times30\arcsec$ subset of the GMOS $i'$ image
  surrounding \qso, marked by 'Q'. Objects mentioned in the text are
  labeled. North and East are indicated by the arrows. The 5 arcsec
  scale bar at bottom is equivalent to 34.9~kpc at the redshift of the
  galaxy group.}
  \label{fig-closeup}
\end{figure}

There are three objects closer to \qso\ than Galaxy \#17. These can be
seen in Fig.~\ref{fig-closeup}, along with some of the group
members. A spectrum of the object west of \qso\ was observed by
\citet{veron90} (their C1). They were unable to measure its redshift,
but its optical and near-infrared colours are similar to their C2 (the
next galaxy to the west), which they found to be at $z=0.21$. It is
therefore likely that this galaxy is foreground to the group at
$z\sim0.66$.

The remaining two nearby objects were not able to be observed in our
GMOS observations (due to the limitations of slit placement). Object
X1 appears to be stellar in morphology, and has colours $i'-K=1.78$ --
not typical of a galaxy at the same redshift as the group. Similarly,
the more distant (from the QLOS) object X3 has a stellar morphology,
and even bluer colours than X1 ($i'-K=1.49$), and thus is likely to be
a star.

The closest object to the quasar, X2, has an extended, non-stellar
morphology, and redder colours than X1 or X3: $i'-K=2.97$, similar to
the colours of the galaxies in the group at $z\sim0.66$. It may thus
be a galaxy in the same system. This putative galaxy would then lie at
a projected distance of $\sim23\,{\rm kpc}$ from the QLOS, making it a
possible candidate as a host of the absorbing system. Its impact
parameter provides a good match to the stronger absorption systems, if
one uses the $W-\rho$ relationship of \citet{lanzetta90}.  A further
point in its favour is its extended, somewhat elongated morphology --
suggestive of a disc galaxy, matching the morphology of many
\mgii-absorbing galaxies.

\subsection{Intra-group gas as an absorber.}

Is the absorbing gas necessarily associated with a galaxy? The fact
that we have a group environment means that the velocity dispersion is
comparable to the rotational velocities of the individual
galaxies. Galaxy interactions are then more likely to strip gas from
galaxies and leave it in the intra-group medium. Such interactions
could date from the formation epoch of the group, or from more
contemporary encounters. An potential example of such an interaction,
based on imaging data, is the pair \#17 and \#18. The velocity
difference between the two ($358\,\kms$), is comparable to the
velocity dispersion of the group, and so is feasible for an
interaction.

\begin{figure}
  \includegraphics{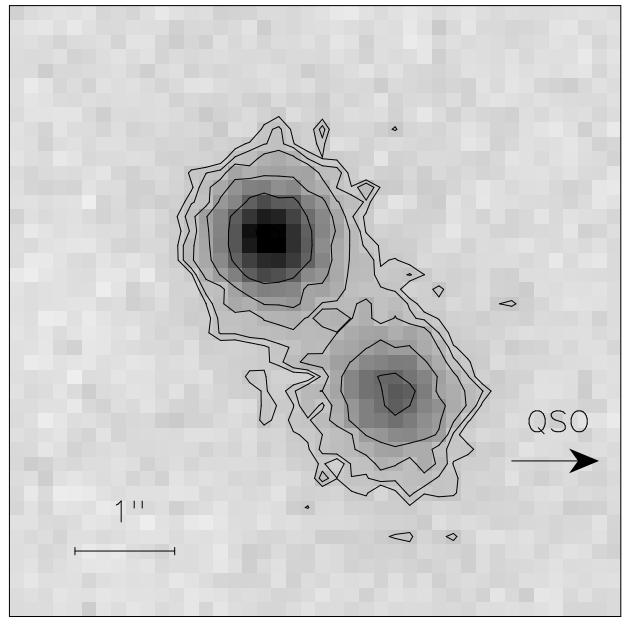}
  \caption{Close-up of Galaxies \#18 (top) and \#17. Orientation is
  the same as in Fig.~\ref{fig-gmos}. Contour levels are at
  $4\sigma, 5\sigma, 7\sigma, 11\sigma, 19\sigma$ above the mean
  background. Field of view is 6.13\,arcsec, and the direction to
  \qso\ is indicated.}
  \label{fig-gal-pair}
\end{figure}

A close-up of this pair of galaxies is shown in
Fig.~\ref{fig-gal-pair}. There is evidence for a connection in flux
between the two, with the flux in the ``bridge'' region reaching
$>7\sigma$ above the background. There is certainly no spectroscopic
sign of any strong interaction taking place, with both galaxies
exhibiting simple passive elliptical spectra. However, sensitive
narrow-band imaging may be able to detect \halpha\ emission from
ejected gas in the vicinity of the galaxies, and would prove a good
test of this interaction hypothesis.

\subsection{Summary and Discussion}

\qso\ is observed to have a strong \mgii\ absorption system at the
same redshift as the group of galaxies detected through our GMOS
spectroscopy. All the nearby galaxies that have spectroscopy are
observed to be ellipticals: each shows a passive elliptical galaxy
spectrum, and none shows evidence for on-going star-formation (for
instance, through the presence of an \oii\ emission line), which would
indicate the presence of large amounts of gas. 

Many \mgii\ absorbers are identified with spiral galaxies
\citep[\eg][]{steidel02}, although this is by no means an exclusive
statement. \citet{steidel94} found that \mgii\ absorption could be
found in galaxies resembling present-day ellipticals as well as
late-type spirals, \citet{jenkins05} found \mgii\ absorption
associated with two S0 galaxies, and recent studies
\citep{churchill05} have found a small fraction of elliptical
absorbers. The latter studies also indicate that the absorption is
also relatively patchy, and so one (or more) of the galaxies could
have isolated areas of absorbing gas that happen to fall across the
QLOS, but are not themselves an important overall part of the
galaxy. This may be similar to the situation observed by
\citet{bowen95}, who failed to find absorption associated with two
elliptical galaxies despite their small impact parameters.

It is possible that there is intra-group gas responsible for the
absorption, although there is little direct evidence for this at this
stage. The best candidate, however, for the absorber is probably
galaxy X2. It is closer to the QLOS, and shows a much less uniform
morphology than the identified galaxies. Ultimately, a spectroscopic
confirmation of its redshift is the only way to be sure -- future
observations will hopefully resolve this question.

\section{The likelihood of gravitational lensing.}
\label{sec-lens}

\qso\ is rather luminous at all frequencies from radio to X-ray. We
would like to know whether it is intrinsically luminous, or has been
magnified through gravitational lensing by some intermediate mass
distribution. Strong lensing can be ruled out, as \qso\ is not
multiply-imaged -- even at VLBI resolution ($\sim1$ mas), it shows a
distinct core-jet structure \citep{stanghellini01}.

Magnification due to gravitational lensing is still possible without
multiple images being detected. There are two contexts in which this
can take place: either a single image is produced, or there are indeed
multiple images but at either faint levels or unresolved
separations. Since VLBI observations place quite strong limits on the
existence of optically-unresolved images, for lensing to be taking
place we would require the production of a single image only.

However, \citet*{keeton05} find that this is possible only when the
lensing is due to massive cluster-sized haloes ($\ga
10^{13.5}M_{\sun}$). Our observations show that the largest mass
concentration along the line-of-sight is merely a galaxy group. The
total luminosity of all galaxies in the group is
$\sim3\times10^{11}L_{\sun}$ (assuming an absolute magnitude of the
Sun of $M_{i'}=4.45$, and taking into account all possible
members, including X2 and the fainter objects in
Fig.~\ref{fig-closeup}). For reasonable mass-to-light ratios, the
group mass will be much less than the threshold given by
\citet{keeton05}. This is supported by the non-detection by
\citet{crawford03} of any extended X-ray emission in {\it Chandra}
images of \qso, indicating the lack of hot cluster gas around or in
the foreground of the quasar. We can thus rule out the possibility
that \qso\ is being magnified by any significant amount. Its large
apparent luminosity is then indeed indicative of its intrinsic power.

\section{Summary}
\label{sec-summary}

We have observed the field surrounding \qso\ with Gemini South + GMOS
in multi-object spectroscopy mode, measuring the redshifts of most of
the nearby galaxies. We find a group of galaxies at $z\sim0.66$,
at a similar velocity to a metal-line absorption system seen in the
quasar's spectrum. The group has five confirmed members, and a further
four fainter galaxies are seen nearby and are possibly
associated. There are a further three galaxies close in redshift but
at large separations on the sky -- while not part of the group, they
are certainly part of the same large-scale structure.

We have also made use of the multiplexing capabilities of GMOS to
measure the redshifts of many other galaxies in the field, and we see
several distinct features in redshift-space foreground to the
group. These observations, in showing the lack of a large cluster in
front of \qso, indicate that the likelihood the quasar is lensed is
small.

While the group as a whole appears to be associated with the
absorption system, the fact that the redshifts of a number of nearby
candidates have not been measured means some doubt remains as to the
identity of the absorber. If the absorber is associated with the
identified galaxies however, this would be a rare example of an \mgii\
absorption system associated with an early-type galaxy.

\section*{Acknowledgments}

We wish to thank the staff at Gemini South for their help in obtaining
the spectroscopic data, particularly acknowledging the help provided
by Michael Ledlow, our contact scientist until his tragic passing. MTW
acknowledges the assistance provided by a NewSouth Global Fellowship
from the University of New South Wales.

This work is based on observations obtained at the Gemini Observatory
under Gemini Program GS-2003B-Q-15. The Gemini Observatory is
operated by the Association of Universities for Research in Astronomy,
Inc., under a cooperative agreement with the NSF on behalf of the
Gemini partnership: the National Science Foundation (United States),
the Particle Physics and Astronomy Research Council (United Kingdom),
the National Research Council (Canada), CONICYT (Chile), the
Australian Research Council (Australia), CNPq (Brazil), and CONICET
(Argentina).

\end{document}